\documentclass[amsmath,amssymb,showpacs,nofootinbib,12pt]{revtex4}

\usepackage{amsmath,amssymb}
\usepackage{graphicx}
\usepackage{epsfig}
\usepackage{cases}

\newcommand{\bdm}{\begin{displaymath}}
\newcommand{\edm}{\end{displaymath}}
\newcommand{\be}{\begin{equation}}
\newcommand{\ee}{\end{equation}}
\newcommand{\ba}{\begin{eqnarray}}
\newcommand{\ea}{\end{eqnarray}}
\newcommand{\bsa}{\begin{subequations}\begin{eqnarray}}
\newcommand{\esa}{\end{eqnarray}\end{subequations}}

\newcommand{\n}{\nonumber}
\newcommand{\lan}{\langle}
\newcommand{\ran}{\rangle}

\newcommand{\defeq}{\mathrel{\mathop:}=}

\newcommand{\ttimes}[1][0pt]{%
    \mathrel{\raisebox{#1}{$\!{\scriptstyle\times}\!$}}%
}

\let\oldin\in \renewcommand{\in}[1][0pt]{%
  \mathrel{\raisebox{#1}{$\!{\scriptstyle\oldin}\!$}}%
}
\let\oldni\ni \renewcommand{\ni}[1][0pt]{%
  \mathrel{\raisebox{#1}{$\!{\scriptstyle\oldni}\!$}}%
}
\let\oldsubseteq\subseteq
\renewcommand{\subseteq}[1][0pt]{%
  \mathrel{\raisebox{#1}{$\!{\scriptstyle\oldsubseteq}\!$}}%
}
\let\oldvarsubsetneq\varsubsetneq
\renewcommand{\varsubsetneq}[1][0pt]{%
  \mathrel{\raisebox{#1}{$\!{\scriptstyle\oldvarsubsetneq}\!$}}%
}
\let\oldsetminus\setminus
\renewcommand{\setminus}[1][0pt]{%
  \mathrel{\raisebox{#1}{$\!{\scriptstyle\oldsetminus}\!$}}%
}
\let\oldbullet\bullet
\renewcommand{\bullet}[1][0pt]{%
  \mathrel{\raisebox{#1}{$\!{\scriptscriptstyle\oldbullet}\!$}}%
}

\begin{document}

\title{Families of Orthogonal Schr\"odinger cat-like-states}
\author{Ludmi{\l}a Praxmeyer\footnote{ {\it{Email address:}}
lpraxm@gmail.com}}
\address{Institute of Photonics Technologies, National Tsing-Hua University,\\
           No. 101, Section 2, Kuang-Fu Road, Hsinchu, Taiwan 30013, R.O.C.}

\begin{abstract}
We analyze condition of orthogonality between optical
Schr{\"o}dinger cat-like-states constructed as superposition of
two coherent states.  We show that the orthogonality condition
leads to quantization of values of a naturally emerging symplectic
form, while values of the corresponding metric form are
continuous. A complete analytical solution of the problem is
presented.
\end{abstract}


\maketitle

\section{Introduction}
A set of nonorthogonal wave functions that naturally appear in the
description of quantum harmonic oscillator was known from the
beginnings of
quantum theory. 
It was firstly mentioned  by Schrodinger's~\cite{Schrodinger} in
1926 and
 analyzed  later in von Neumann's  {\it Mathematische Grundlagen der
Quantenmechanik} \cite{vonNeumann}. The name `coherent states' was
proposed by Glauber in the context  of description of coherent
laser beams  in 1963~\cite{Glauber}. Since then, formalism of
coherent states often serves as a
 language of quantum optics -- especially in its phase
space representation. Not without reason: as the `most classical
from quantum states'  coherent states match 
classical intuitions, when at the same time superposition of
coherent states are nonclassical enough to reveal purely quantum
effects. As a typical example usually serves 
a superposition of two coherent states, a so-called {\sl
Schr\"odinger-cat-like state}.

Superpositions of coherent states have been studied in the
contexts of quantum error correction
\cite{Cochrane}, quantum teleportation
\cite{Enk}, and quantum computation
\cite{Kim,Glancy}. The majority of these applications exploit the
fact that a cat-like state  split on a beamsplitter produces an
entangled state. There are many theoretical proposals of
generation of such superposition and a number of experimental
realizations~\cite{ourj,ourj2,Takahashi}. I believe that many from
 proposals mentioned above could benefit from the use of additional
fact that, unlike single coherent states, their superpositions
form families of orthogonal states. This paper presents an
analysis showing that the scalar product between superpositions of
coherent states can be exactly zero -- not just reaches `close to
zero' values, as it is the case when single coherent states are
used, e.g., as logical qubits.

Coherent states minimize uncertainty relation; they are connected
to the eigenvectors of a quantum harmonic oscillator,
$|m\ran_{\!{}_{osc}}$, by the formula
\ba
|\alpha\ran=e^\frac{-|\alpha|^2}{2}\sum_{m=0}^\infty
\frac{\alpha^m}{\sqrt{m!}} |m\ran_{\!{}_{osc}},\label{cd}
\ea
and they form an over-complete (but not orthogonal) basis. Average
number of photons\footnote{Because we consider coherent states in
the context of quantum optics we talk about average number of
photons rather then average number of excitations.} in (\ref{cd})
is equal to $|\alpha|^2$.
 Although two coherent states are never orthogonal to each other, the scalar
product between them vanishes exponentially with distance  $
\big|\lan \beta|\alpha \ran\big|^2\!=\exp({-|\alpha-\beta|^2})$,
 which for large enough values of $|\alpha| $ allows to treat a sum
\ba
{\mathcal{K}}_{\!{}_\varphi}\!(\alpha):=|\alpha\ran+e^{i\varphi}|-\alpha\ran\,,\label{e1}
\ea
as a close optical analogue of the superposition of
macroscopically distinguishable states from the Schr\"odinger's
{\sl{gedanken}} experiment
\cite{Schrodinger_cat}. For small values of $|\alpha|$ (and
significant overlap between the states) such superpositions are
known as {\sl Schr\"odinger kitten}. Because of an omitted
normalization factor,
we shall refer to ${\mathcal{K}}_{\!{}_\varphi}(\alpha)$
  as to vector rather then a state.
For $\varphi=0$ and $\varphi=\pi$, vector
${\mathcal{K}}_{\!{}_\varphi}(\alpha)$ corresponds respectively to
even and odd coherent states introduced in
\cite{Hillery}. Note, also,  that the phase in~(\ref{e1}) is not equivalent
the one studied in~\cite{Kien} where superpositions of the form  $|
\alpha e^{i\phi}\ran+|\alpha e^{-i\phi}\ran$ were considered.

The Wigner function
\cite{Wigner} of a cat-like state is often used to illustrate  how decoherence
affects quantum superpositions \cite{BuzekKnight,Zurek_decoh}. In
this phase space representation it is clearly seen that addition
of a noise to the system destroys the interference terms, while
Gaussian peaks corresponding to $|\alpha\ran$ or $|-\alpha\ran$
remain unaffected. It was shown, that the coherent states 
are especially robust to decoherence, thus, can serve as the
`pointer' states
\cite{Zurek_decoh}.
 More information about mathematical properties of coherent states
and their generalizations can be found in \cite{Perelomov,Gazeau}.
A short list of symbols used in this paper is presented in
Appendix~A.

\section{Superposition of coherent states}
\subsection{A coherent state is orthogonal to ...}\label{s2} We have
emphasized that coherent states are not orthogonal to each other.
It does not mean that also superpositions of coherent states
always have non-zero scalar product. The most obvious example of a
vanishing overlap between Schr\"odinger-cat-like states  is
obtained via change of a relative phase: vector
${\mathcal{K}}_{\pi}(\alpha)$ is orthogonal to
${\mathcal{K}}_{\scriptscriptstyle{0}}(\alpha)$. This fact is
 easily proved,  one just have to realize the difference in the parity of the
corresponding states. Substituting~(\ref{cd}) into
${\mathcal{K}}_{\scriptscriptstyle{0}}(\alpha)$ and
${\mathcal{K}}_{\pi}(\alpha)$ one sees that the  former consists
only of even number states, while the later only of odd number
states \cite{Hillery}. The same argument proves that
${\mathcal{K}}_{\pi}(\alpha)$ and
${\mathcal{K}}_{\scriptscriptstyle{0}}(\beta)$ are always
orthogonal and that the vacuum state $|0\ran$ is orthogonal to
${\mathcal{K}}_{\!\pi}(\alpha)$ for any $\alpha$. Another example
is a class of  vectors orthogonal to
${\mathcal{K}}_{\scriptscriptstyle{0}}(d)$, $ d\in[1pt]
\mathbb{R}^\times,$ defined as
\ba
{\mathcal{J}}\!(d,\delta_k):=|d+i\delta_k\ran+|-d+i\delta_k\ran\,,\n
\ea
where $\delta_k=\pi(2k+1)/(2d)$ and $k\in[1pt]\mathbb{Z}$. Each
value of $\delta_k$ corresponds to a shift in momentum that makes
${\mathcal{K}}_{\scriptscriptstyle{0}}(d_{})$ and
${\mathcal{J}}\!(d,\delta_k)$ orthogonal to each other
\cite{frogTorun}.

One might wonder if a superposition
${\mathcal{K}}_\varphi(\alpha)$ could be orthogonal to a single
coherent state different then $|0\ran$. The answer is yes: for
example an overlap of vector $|\alpha\ran$ for any  set but
nonzero real $\alpha$, and
${\mathcal{K}}_{\scriptscriptstyle{0}}(\beta_n)$ for
\ba
\beta_n=i\pi(n+{1}/{2})/\alpha,
\label{w1}
\ea
vanishes for all natural $n$, i.e.,
\ba
\lan\alpha|{\mathcal{K}}_{\scriptscriptstyle{0}}\!
{\textstyle\big(i\pi(n+{1}/{2})/\alpha\big)}\ran=0 \qquad
\mathrm{for}\;
\alpha\in[1pt]\mathbb{R}^{\times},\; n\in[1pt]\mathbb{N}. \n
\ea
For a set value of $n$, the smaller $|\alpha|$ the further apart
are $|\beta_n\ran$ and $|-\beta_n\ran$; for a set $\alpha$, the
amplitude $|\beta_n|$ increases with $n$.

\begin{figure}[h]
\noindent\rule{15.4cm}{0.4pt}
 \begin{center}
{\footnotesize{a)}}$\!\!$
\includegraphics[width=4cm]{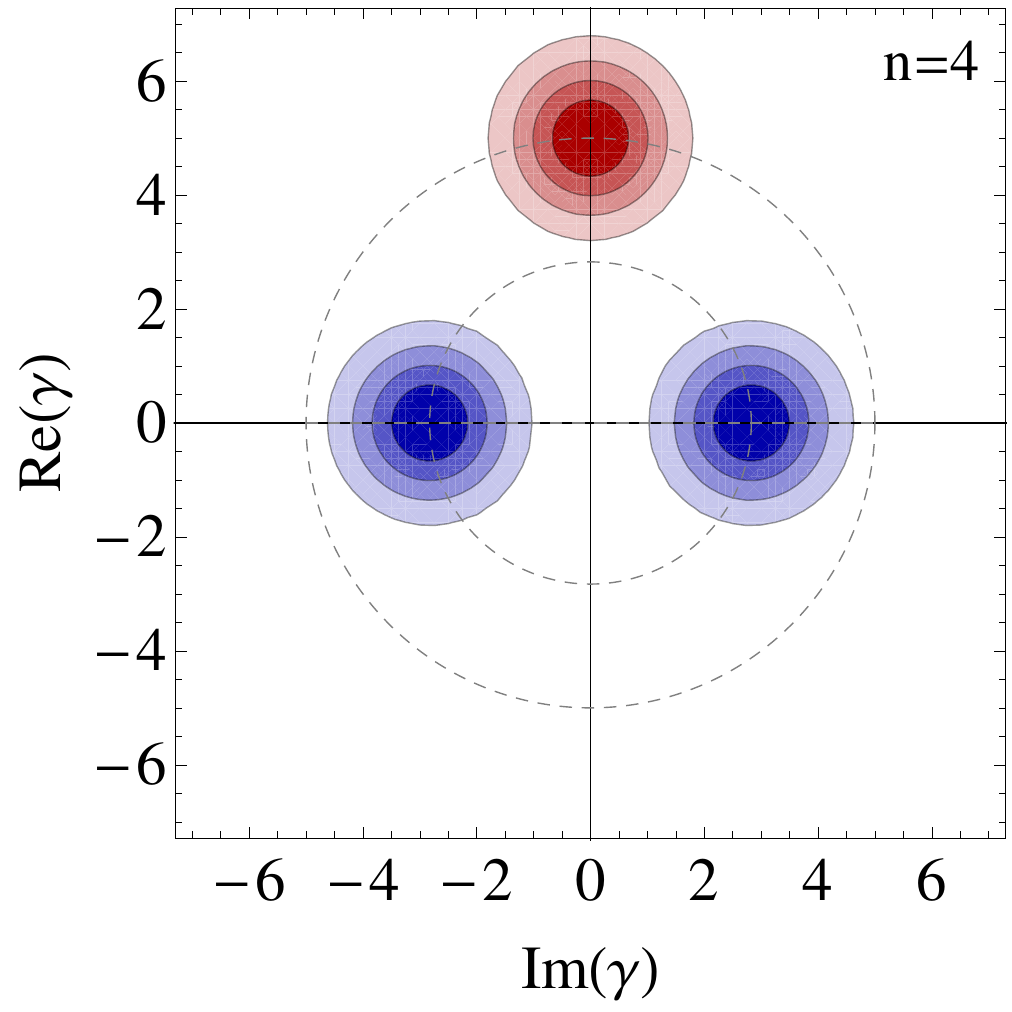}$\;$   
{\footnotesize{b)}}$\!\!$
\includegraphics[height=4cm]{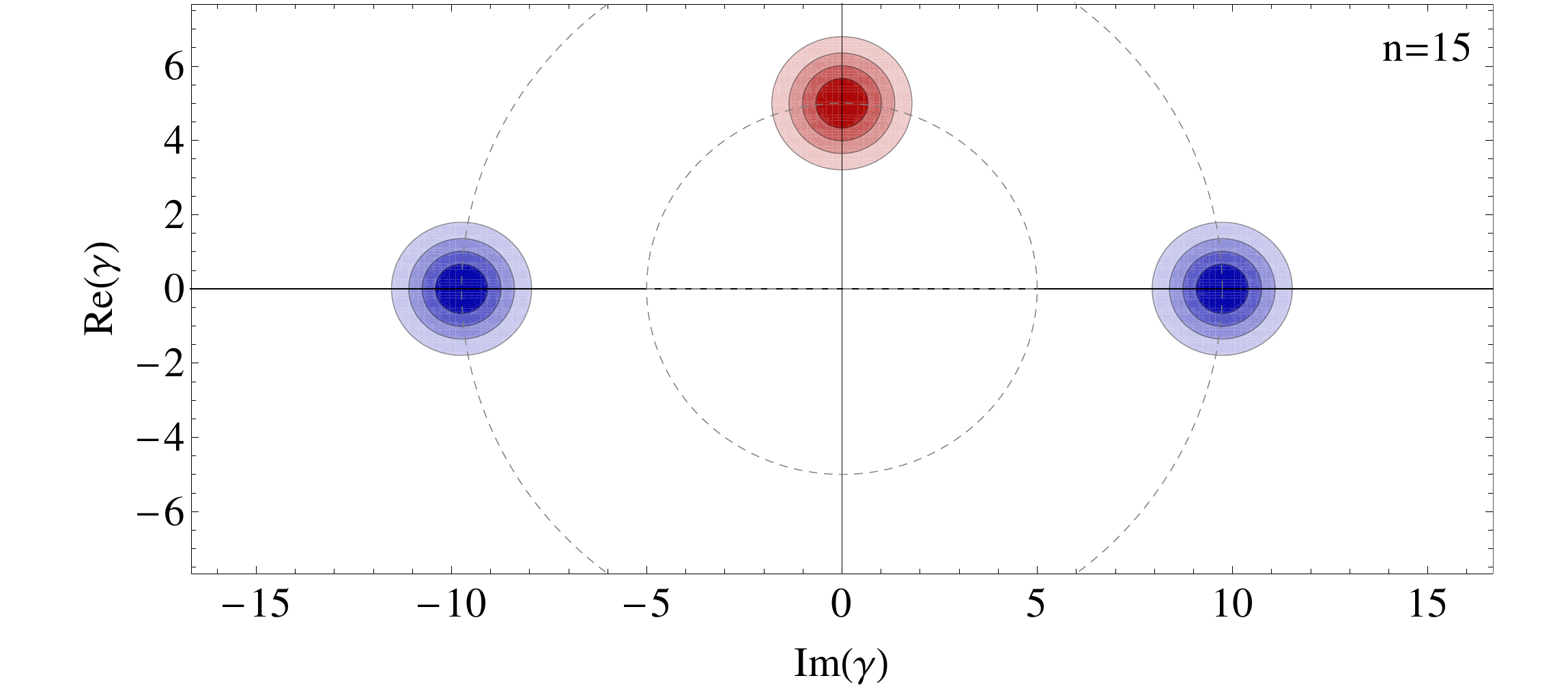}
\end{center}
\vspace*{-.5cm}
\caption{ Phase space Husimi representation of coherent state
$|5\ran$ (plotted in red) and orthogonal to it cat-like
superposition ${\mathcal{K}}_{\scriptscriptstyle{0}}(\beta_n)$
(plotted in blue). Amplitudes of orthogonal states take on
discrete values and increase with n. Blue circles in plots a) and
b) correspond to vectors
 ${\mathcal{K}}_{\scriptscriptstyle{0}}(i \pi (0.9))$ and
  ${\mathcal{K}}_{\scriptscriptstyle{0}}(i\pi (3.1))$ obtained from (\ref{w1})
  for $n=4$   or  n=15, respectively.             }
\label{fig0}
\noindent\rule{15.4cm}{0.4pt}
\end{figure}

To illustrate examples used, we will  either plot the
corresponding Husimi functions~\cite{Husimi} or just represent a
coherent state $|\alpha\ran$ as a circle of radii $1/\sqrt{2}\,$
centered at point $(\mathrm{Re}(\alpha),\mathrm{Im}(\alpha))$. For
a given density matrix~$\hat\varrho$, Husimi function is
defined as
\ba
\mathrm{Q}_{\hat\varrho}(\gamma)=Tr[\hat\varrho
|\gamma\ran\lan\gamma|].\label{hus}
\ea
Figure~\ref{fig0} shows  Husimi functions corresponding to
cat-like
 superpositions orthogonal to coherent state
 $|5\ran$.
The first example, depicted in Fig.~\ref{fig0}.a,   corresponds to
 $n=4$ and vector~${\mathcal{K}}_{\scriptscriptstyle{0}}( i \pi (0.9) )$.
 The second,  Fig.~\ref{fig0}.b,
 was calculated
 for $n=15$ and corresponds
 to vector~${\mathcal{K}}_{\scriptscriptstyle{0}}(i\pi(3.1))$.
In both figures,   Husimi function of state
 $|5\ran$ is plotted in red, and Husimi functions of the
 orthogonal vectors  ${\mathcal{K}}_{\scriptscriptstyle{0}}(\beta_n)$
are  plotted in blue.
 Dashed circles of radii $5$ and
$ |\beta_n|$ denote phase-space trajectories of free evolution of
coherent states $|5 \ran$, $|\beta_n\ran$. Note, that in respect
to typical notation position ($\sqrt{2}
\mathrm{Re}(\gamma)$)  and momentum  ($\sqrt{2}
\mathrm{Im}(\gamma)$) axes in these plots are exchanged.

\subsection{Even and odd coherent states}
\subsubsection{Orthogonality between even cats} 
\begin{figure}[h]
\noindent\rule{15.4cm}{0.4pt}
 \begin{center}
{\footnotesize{a)}}%
\rotatebox{90}{\hspace*{2cm}\it\scriptsize
$\mathrm{Im}(\gamma)$}$\!\!\!$
\includegraphics[width=4.5cm]{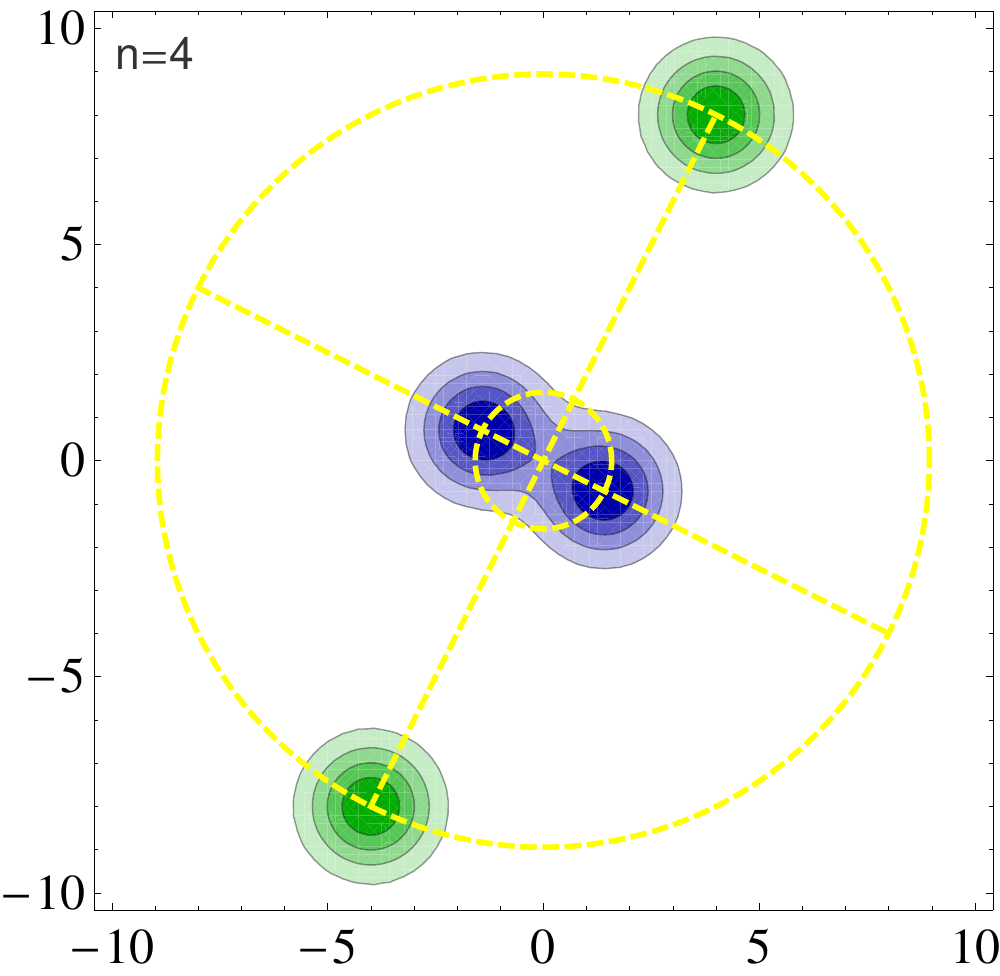}$\;$   
{\footnotesize{b)}}\includegraphics[width=4.5cm]{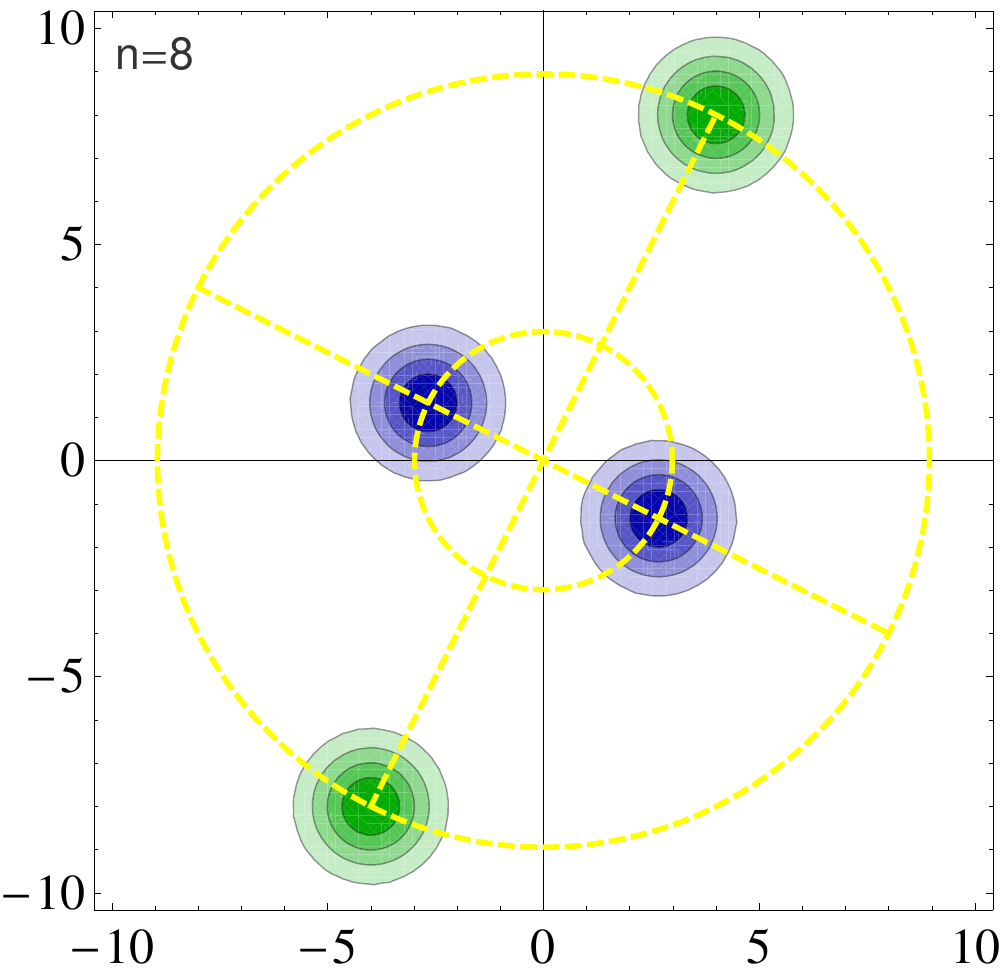}$\;$
{\footnotesize{c)}}\includegraphics[width=4.5cm]{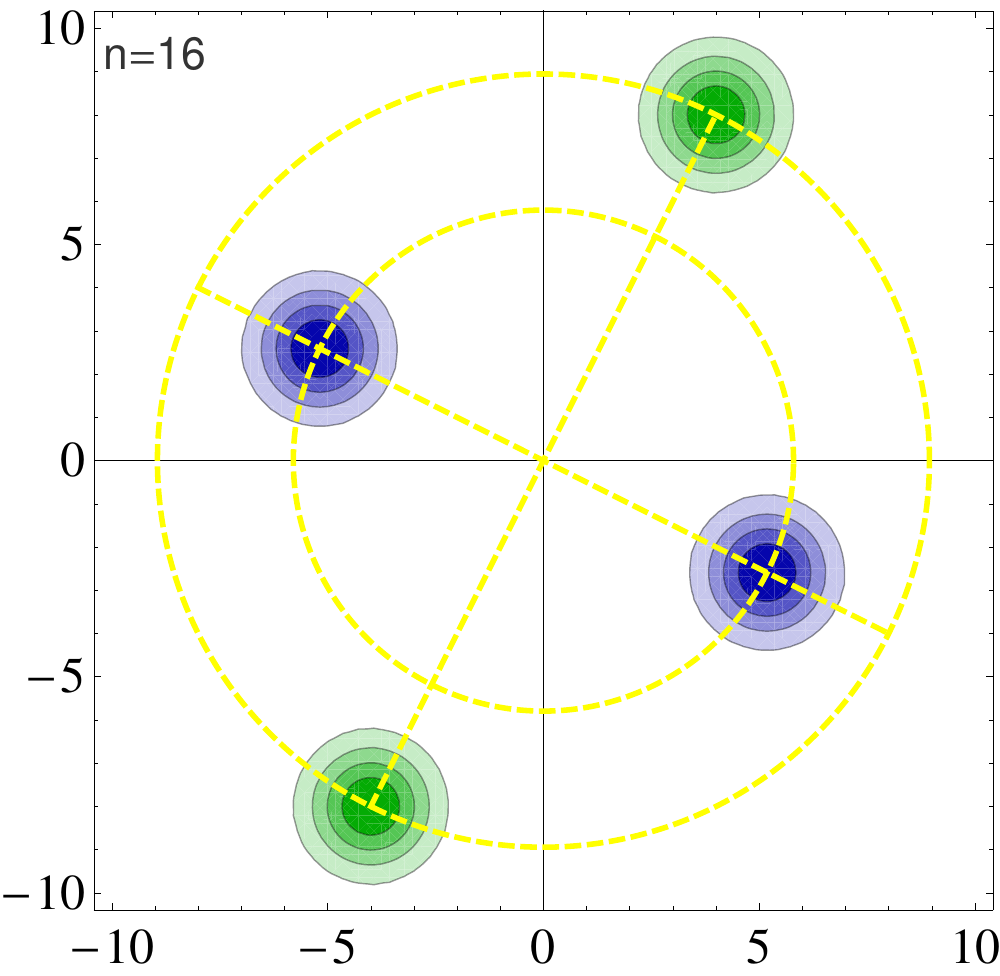}
\end{center}
\vspace*{-.5cm}
\hspace*{1cm}{\it\scriptsize $\mathrm{Re}(\gamma)$}\hspace*{4.4cm}{\it\scriptsize $\mathrm{Re}(\gamma)$}
\hspace*{4.cm}{\it\scriptsize $\mathrm{Re}(\gamma)$}
\caption{ Husimi functions corresponding to the pairs of orthogonal Schrodinger-like-cat superpositions.
Initial cat  {\mbox{${\mathcal{K}}_{\scriptscriptstyle{0}}\!(4+i8)
$}} is depicted in green. Plotted in blue are Husimi functions
corresponding to
${\mathcal{K}}_{\scriptscriptstyle{0}}\!(\beta_n)$ from
(\ref{fakt1}), for a) n=4; b) n=8; c) n=16, respectively. }
\label{fig1}
\noindent\rule{15.4cm}{0.4pt}
\end{figure}

Consider a
superposition of two coherent states of form $
{\mathcal{K}}_{\scriptscriptstyle{0}}\!(\alpha)$, Eq.~(\ref{e1}).
It can be proved that: 

\noindent
{\bf Fact 1} For  any $\alpha,\beta\in[1.5pt]
\mathbb{C}^\times$, the following
conditions are
equivalent:\\
\noindent
$1^\circ\quad$ $\lan{\mathcal{K}}_{\scriptscriptstyle{0}}(\beta)|
{\mathcal{K}}_{\scriptscriptstyle{0}}(\alpha)\ran=0\,.$\\
$2^\circ\quad$     
There exists $n\in[1pt]\mathbb{N}$, such that
\ba
{\mathcal{K}}_{\scriptscriptstyle{0}}(\beta)={\mathcal{K}}_{\scriptscriptstyle{0}}(\beta_n):=
 |-i\pi(2n+1)/(2\alpha^*)\ran + | i \pi(2n+1)
 /(2\alpha^*)\ran\,.\label{fakt1}
 \ea
It is seen, that for a given non-zero $\alpha$ there exists a
whole class of solutions of $1^\circ\,$ parametrized by natural
number $n$, each rotated in
phase space by $\pi/2$ in respect to 
${\mathcal{K}}_{\scriptscriptstyle{0}}(\alpha)$. Figure~\ref{fig1}
shows examples of Husimi functions calculated for pairs of
orthogonal vectors ${\mathcal{K}}_{\scriptscriptstyle{0}}(\alpha)$
(plotted in green) and $
{\mathcal{K}}_{\scriptscriptstyle{0}}(\beta_n)$ (plotted in blue),
for parameters $\alpha=4+i8$ and $\beta_n$
corresponding to $n=4$, {\mbox{$n=8$,}} $n=16$, respectively.
Comparison between Figs.~\ref{fig1}~a), b) and c) reveals how
 change of $n$ changes distance between states forming
 superposition (\ref{fakt1}).
 In general, separation $2|\beta_n|$
 is different then $2|\alpha|$
 for every value of parameter $n$. Below we consider a very special
 case when the distances between states forming orthogonal
 superpositions ${\mathcal{K}}_{\scriptscriptstyle{0}}(\alpha)$
  and ${\mathcal{K}}_{\scriptscriptstyle{0}}(\beta)$ are equal.

\noindent
{\bf Fact 2} For $\alpha,\beta\in[1.5pt] \mathbb{C}^\times$ such
that
 $ \lan{\mathcal{K}}_{\scriptscriptstyle{0}}(\alpha)|
       {\mathcal{K}}_{\scriptscriptstyle{0}}(\beta)\ran=0$
the following conditions are equivalent:\\
$1^\circ\quad$ $|\alpha|=|\beta|$\,.\\
$2^\circ\quad$ There exists $n\in[1pt]\mathbb{N}$ such that
$|\alpha|=\sqrt{(n+1/2)\pi} \;.$

\noindent
Example of two even cat superpositions with equal average number
of photons is presented in Fig.~\ref{fig6}.a). Dashed circles show
the only possible values of~$\alpha$ satisfying condition
$2^\circ$ from Fact 2. Areas of bands between the subsequent
circles are equal to~$\pi^2$.

\subsubsection{Orthogonality between odd cats}\label{odd}

Similarly, one might consider  two coherent states forming an
`odd' cat-like superposition $
{\mathcal{K}}_{\scriptstyle{\pi}}\!(\alpha')$ and prove the facts
listed below:

\noindent
{\bf Fact 3} For  any $\alpha',\beta'\in[1.5pt]
\mathbb{C}^\times$, the following conditions are equivalent:\\
\noindent
$1^\circ\quad$ $\lan{\mathcal{K}}_{\scriptstyle{\pi}}(\beta')|
{\mathcal{K}}_{\scriptstyle{\pi}}(\alpha')\ran=0\,.$\\
$2^\circ\quad$     
There exists $n\in[1pt]\mathbb{N}^\times$, such that
${\mathcal{K}}_{\scriptstyle{\pi}}(\beta')=
{\mathcal{K}}_{\scriptstyle{\pi}}(\beta'_n):=  |-i n\pi/\alpha'^*
\ran - |  i n\pi/\alpha'^*\ran\,. $\\
(Note that, for $n=0$, the orthogonality condition holds, although
it is reduced to a trivial case.)

\noindent
{\bf Fact 4} For $\alpha',\beta'\in[1.5pt] \mathbb{C}^\times$ such
that  $ \lan{\mathcal{K}}_{\scriptstyle{\pi}}(\alpha')|
       {\mathcal{K}}_{\scriptstyle{\pi}}(\beta')\ran=0$
the following conditions are equivalent:\\
$1^\circ\quad$ $|\alpha'|=|\beta'|$\,.\\
$2^\circ\quad$ There exists $n\in[1pt]\mathbb{N}^\times$ such that
$ {\textstyle |\alpha'|=\sqrt{n\pi} } \;.$\\

\begin{figure}[h]
\noindent\rule{15.4cm}{0.4pt}\\

\begin{minipage}[b]{0.45\textwidth}
$\;$\\

\begin{center}
\includegraphics[width=5.5cm]{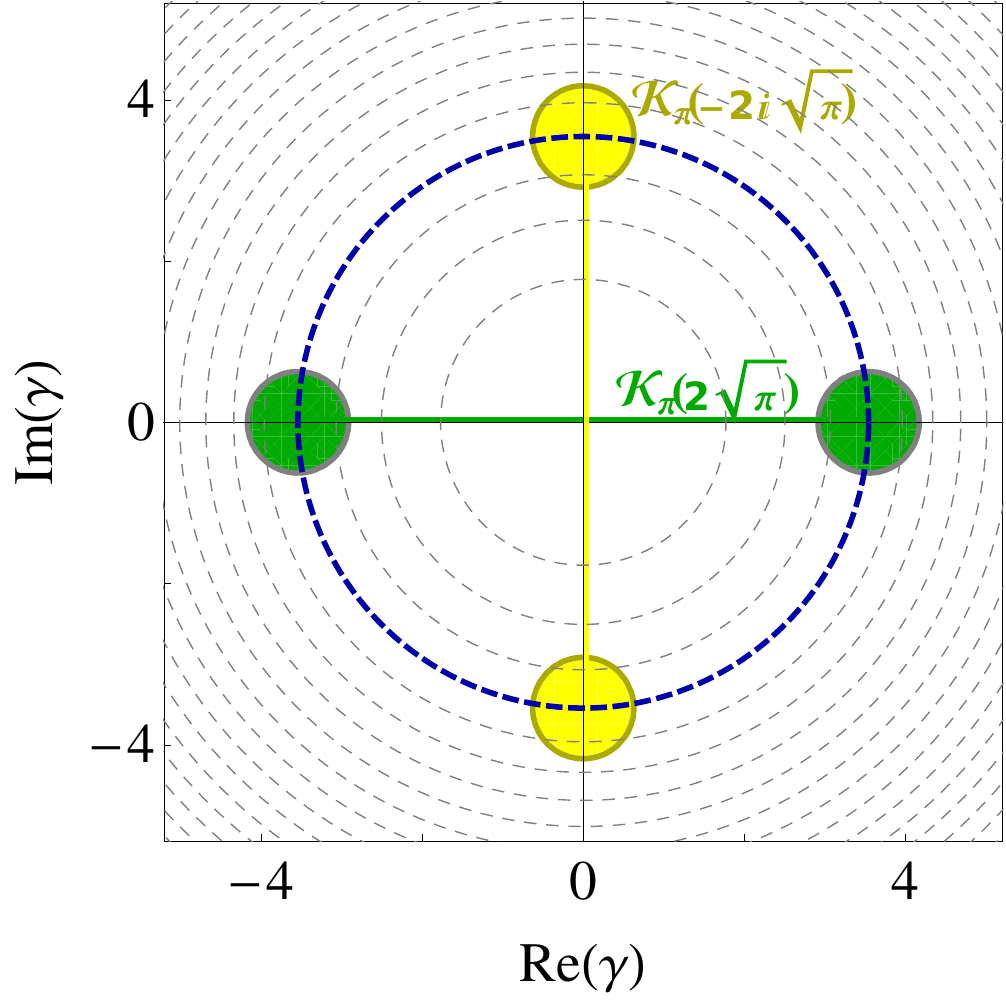}  
\end{center}
\end{minipage}
\begin{minipage}[t]{0.5\textwidth}
\vspace*{-6.cm}
\caption{ Phase space representation of a pair of orthogonal odd
cats with the same average number of photons. Superposition
{\mbox{${\mathcal{K}}_{\pi}(\alpha')\!=|\sqrt{4\pi}\ran-|-\sqrt{4\pi}\ran\,
$}} is plotted in green and the orthogonal superposition
{\mbox{${\mathcal{K}}_{\pi}(\beta')\!=|-i\sqrt{4\pi}\ran-|i\sqrt{4\pi}\ran\,
$}}  in yellow. Dashed lines denote circles of radii
$r=\sqrt{n\pi}$ showing the only possible values such that
$|\beta'|=|\alpha'|$ and $
\lan{\mathcal{K}}_{\scriptstyle{\pi}}\!(\alpha')
|{\mathcal{K}}_{\scriptstyle{\pi}}\!(\beta')\ran=0$.}\label{fig2}
\end{minipage}
\noindent\rule{15.4cm}{0.4pt}
\end{figure}

Figure \ref{fig2} shows a phase space representation of pair of
orthogonal  vectors ${\mathcal{K}}_{\scriptstyle{\pi}}(\alpha'),
{\mathcal{K}}_{\scriptstyle{\pi}}(\beta')$  fulfilling condition
$|\alpha'|=|\beta'|$. Black dashed lines denote other circles of
radii  of form $r_n=\sqrt{n\pi}.$ As before, areas of bands
between the closest-neighbors circles are equal to $\pi^2$. There
is a  factor of $\sqrt{\pi}$ difference between these radii and
those separating the subsequent Planck-Bohr-Sommerfeld bands
\cite{Kien} corresponding to number states in the semiclassical
limit.

\section{Orthogonality between cat-like states -- arbitrary relative phase}
Let us consider a more general superpositions obtained
 for non-zero $\alpha$, $\beta$ and
arbitrary relative phases $\varphi_1,\varphi_2 \in[1.pt]
[0,2\pi[$. It can be shown that a condition
\ba
\big|\lan{\mathcal{K}}_{\varphi_{\!{}_2}}\!(\beta)|{\mathcal{K}}_{\varphi_{\!{}_1}}\!(\alpha)\ran\big|^2=0
\;\label{c5}
\ea
is equivalent to equation
\ba
  e^{2\mathrm{Re}(\alpha\beta^*)}\!\cos^2\!\!\big({\textstyle{\frac{\varphi_{\!{}_2}-\varphi_{\!{}_1}}{2}}}\big)
   \! +\!
    e^{-2\mathrm{Re}(\alpha\beta^*) }\!\cos^2\!\!\big({\textstyle{\frac{\varphi_{\!{}_2}+\varphi_{\!{}_1}}{2}}}\big)
 \!=\n\\
 =-2
\cos\!\big({\textstyle{\frac{\varphi_{\!{}_2}+\varphi_{\!{}_1}}{2} }}\big)
\cos\!\big({\textstyle{\frac{\varphi_{\!{}_2}-\varphi_{\!{}_1}}{2} }}\big)
\cos\!\big[2\,\mathrm{Im}(\alpha\beta^*) \big]
\label{scal}
   . \ea
In the paragraphs that follow,  analysis od solutions of
(\ref{scal}) is presented and dependance between parameters
$\alpha,\,\beta,\,\varphi_{\!{}_1},\,\varphi_{\!{}_2}$ examined.

   \subsection{Case when $\cos({\scriptstyle{\frac{\varphi_{\!{}_2}+\varphi_{\!{}_1}}{2} }})
\cos({\scriptstyle{\frac{\varphi_{\!{}_2}-\varphi_{\!{}_1}}{2}
}})= 0\,$}

\begin{figure}[h]
\noindent\rule{15.4cm}{0.4pt}
 \begin{center}
{\footnotesize{a)}}\includegraphics[width=4.1cm]{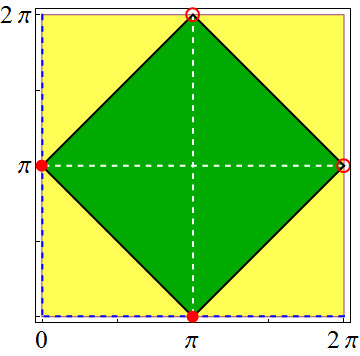}   
$\qquad\qquad$
{\footnotesize{b)}}\includegraphics[width=4.8cm,angle=0]{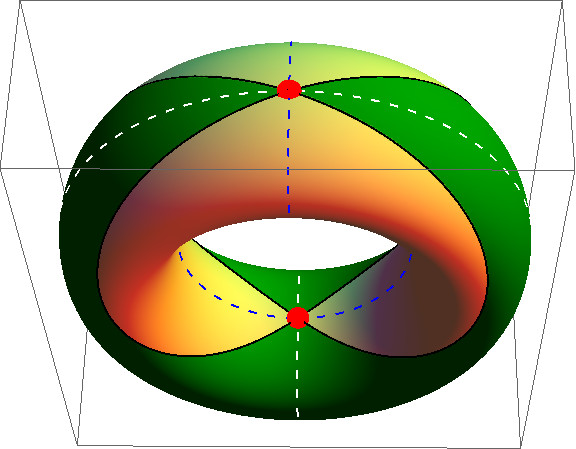}   
\end{center}
\caption{(a) Map of $(\varphi_{\!{}_1},\varphi_{\!{}_2})$ that organizes solutions of
(\ref{scal}). Its periodic boundary conditions make it equivalent
to a torus (b) showing that the green and yellow areas have the
same topology. Red dots correspond to the case of odd and even
cats which are orthogonal to each other for all values of
$\alpha$, $\beta$. Dashed white and dashed blue lines correspond
to cases when $\mathrm{Re}(\alpha
\beta^*)=0$ and one of the phases is equal to 0 or $\pi$ while the
second is arbitrary. Values of $\varphi_{\!{}_1},\varphi_{\!{}_2}$
for which
$\cos({\scriptstyle{\frac{\varphi_{\!{}_2}-\varphi_{\!{}_1}}{2}
}})/
\cos({\scriptstyle{\frac{\varphi_{\!{}_2}+\varphi_{\!{}_1}}{2}
}})<0$ are plotted as a green  inscribed square. Its black edges
(open segments) denote the values for which only one of these
cosines vanishes, and there are no solutions of (\ref{scal}). The
remaining yellow triangles denote these
$\varphi_{\!{}_1},\varphi_{\!{}_2}$ for which
$\cos({\scriptstyle{\frac{\varphi_{\!{}_2}-\varphi_{\!{}_1}}{2}
}})/
\cos({\scriptstyle{\frac{\varphi_{\!{}_2}+\varphi_{\!{}_1}}{2}
}})>0$. For angles forming open green and yellow areas there are
always unambiguous solutions of (\ref{scal}). }
\label{rysTorus}
\noindent\rule{15.4cm}{0.4pt}\\
\end{figure}

\noindent
Note that when both $
\cos({\scriptstyle{\frac{\varphi_{\!{}_2}+\varphi_{\!{}_1}}{2}
}})=0 $ and                            
$\cos({\scriptstyle{\frac{\varphi_{\!{}_2}-\varphi_{\!{}_1}}{2}
}})=0, $ (\ref{scal}) forms an identity.   As a result,  the
scalar product (\ref{c5}) vanishes for any $\alpha$ and $\beta$.
On a map presented  in Fig.~\ref{rysTorus}.a,  phases
$(\varphi_{\!{}_1},\varphi_{\!{}_2})\in[1pt]
\{(0,\pi),(\pi,0) \} $ corresponding to this case  are illustrated as red circles. It is
an example of orthogonality between odd and even cats, that occurs
regardless of $\alpha$, $\beta$, which was already described in
the beginning of Section~\ref{s2}.

Quite the opposite is the case when
$\cos({\scriptstyle{\frac{\varphi_{\!{}_2}+\varphi_{\!{}_1}}{2}
}})
\cos({\scriptstyle{\frac{\varphi_{\!{}_2}-\varphi_{\!{}_1}}{2}
}})= 0\, $ but either
$\cos({\scriptstyle{\frac{\varphi_{\!{}_2}-\varphi_{\!{}_1}}{2}
}})\neq 0$ or
$\cos({\scriptstyle{\frac{\varphi_{\!{}_2}+\varphi_{\!{}_1}}{2}
}})\neq 0$. Then,  there are no solutions of (\ref{scal}). In
Fig.~\ref{rysTorus}.a, phases corresponding to this condition are
plotted as black open edges of an
 inscribed green square. These lines divide map $(\varphi_{\!{}_1},\varphi_{\!{}_2})$
into two areas: a green square where
{\mbox{$\cos({\scriptstyle{\frac{\varphi_{\!{}_2}-\varphi_{\!{}_1}}{2}
}})/
\cos({\scriptstyle{\frac{\varphi_{\!{}_2}+\varphi_{\!{}_1}}{2}
}})<0$}}  and a remaining yellow area where
{\mbox{$\cos({\scriptstyle{\frac{\varphi_{\!{}_2}-\varphi_{\!{}_1}}{2}
}})/
\cos({\scriptstyle{\frac{\varphi_{\!{}_2}+\varphi_{\!{}_1}}{2}
}})>0$.}} Because of periodic boundary conditions both surfaces
have exactly the same topology, as it is seen on
Fig.~\ref{rysTorus}.b.

\subsection{Case when $\cos({\scriptstyle{\frac{\varphi_{\!{}_2}+\varphi_{\!{}_1}}{2} }})
\cos({\scriptstyle{\frac{\varphi_{\!{}_2}-\varphi_{\!{}_1}}{2}
}})\neq 0\,$}

\noindent
To find solutions of (\ref{scal}) in a 
nonsingular case of
$\cos({\scriptstyle{\frac{\varphi_{\!{}_2}+\varphi_{\!{}_1}}{2} }})
\cos({\scriptstyle{\frac{\varphi_{\!{}_2}-\varphi_{\!{}_1}}{2}
}})\neq 0\, $, it is convenient to rewrite this equation as
\ba
-2
\cos\!\big[2\,\mathrm{Im}(\alpha\beta^*) \big]=
 \exp\!\big[2 \mathrm{Re}(\alpha\beta^{\ast})\big]
 \cos\!\big({\textstyle\frac{\varphi_{\!{}_1} -  \varphi_{\!{}_2}}{2}}\big)
 /\cos\!\big({\textstyle\frac{\varphi_{\!{}_1} +
 \varphi_{\!{}_2}}{2}}\big)+\n\\
 +
 \exp\!\big[\!-\!2 \mathrm{Re}(\alpha\beta^{\ast})\big]
 \cos\!\big({\textstyle\frac{\varphi_{\!{}_1} + \varphi_{\!{}_2}}{2}}\big)
/\cos\!\big({\textstyle\frac{\varphi_{\!{}_1} -
 \varphi_{\!{}_2}}{2}}\big).
\label{scala}
   \ea
Values of the left-hand side of (\ref{scala}) form closed segment
$[-2,2]$. The right-hand side took the form of function
$f(z)=z+1/z,$ and its values belong to $\,[2,\infty[\,$ or
$\,]-\infty,-2]$ for positive and negative $z$, respectively. It
is easy to check that $f(z)=-2$ iff $z=-1$, and $f(z)=2$ iff
$z=1$. Combination of these facts leads to conclusion that
(\ref{scala}) is satisfied only when its both sides are
simultaneously equal to plus or minus~2. Thus, for (\ref{c5}) to
hold\footnote{in the case when
$\cos({\scriptstyle{\frac{\varphi_{\!{}_2}+\varphi_{\!{}_1}}{2}
}})\cos({\scriptstyle{\frac{\varphi_{\!{}_2}-\varphi_{\!{}_1}}{2}
}})\neq 0\,$} one from the following conditions has to be
satisfied:
 \begin{subequations}
    \begin{numcases}{}
      \exp\!\big[2 \mathrm{Re}(\alpha\beta^{\ast})\big]
  =  - \cos\!\big({\textstyle\frac{\varphi_{\!{}_1} + \varphi_{\!{}_2}}{2}}\big)
  / \cos\!\big({\textstyle\frac{\varphi_{\!{}_1} -  \varphi_{\!{}_2}}{2}}\big)
   \label{10A} \\
     \mathrm{Im}(\alpha\beta^{\ast})=k\pi
    \qquad\qquad\qquad\qquad\qquad\qquad\qquad{\mbox{for}}
     \quad k\in[1.pt]
\mathbb{Z}, \label{10B}
    \end{numcases}
  \end{subequations}
or
\begin{subequations}
    \begin{numcases}{}
      \exp\!\big[2 \mathrm{Re}(\alpha\beta^{\ast})\big]
  =  \cos\!\big({\textstyle\frac{\varphi_{\!{}_1} + \varphi_{\!{}_2}}{2}}\big)
  / \cos\!\big({\textstyle\frac{\varphi_{\!{}_1} -  \varphi_{\!{}_2}}{2}}\big)\label{11A} \\
     \mathrm{Im}(\alpha\beta^{\ast})=(2k+1)\pi/2 \;
      \qquad\qquad\qquad\qquad\qquad{\mbox{for}}\quad k\in[1.pt]
\mathbb{Z}. \label{11B}
    \end{numcases}
  \end{subequations}
\noindent
Note, that requirements imposed on
$\mathrm{Im}(\alpha\beta^{\ast}) $ and
$\mathrm{Re}(\alpha\beta^{\ast}) $ are separated:  phases
$\varphi_{\!{}_1}$, $\varphi_{\!{}_2}$ define
$\mathrm{Re}(\alpha\beta^{\ast}) $ unambiguously,  but are
independent from the quantization conditions imposed on
$\mathrm{Im}(\alpha\beta^{\ast}) $. It is also worth noting, that
the quantization conditions are imposed only on a naturally
emerging symplectic form, $\mathrm{Im}(\alpha\beta^{\ast}) $,
while values of the corresponding metric form,
$\mathrm{Re}(\alpha\beta^{\ast}), $ can change continuously. For
more details, see Appendix B.

Let us assume that $\alpha
\neq 0$ and $\mathrm{Re}(\alpha\beta^*)=\omega $.  From (\ref{10B}) or (\ref{11B})
we obtain
\ba
 \beta = (\omega - i k\pi)/\alpha^{*} \qquad \mathrm{or} \qquad \beta = (\omega - i(k
+ 1/2)\pi)/\alpha^{*},\qquad k\in[1pt] \mathbb{Z} \label{e12}
\ea
 respectively. It is seen that for every real $\omega$ and non-zero $\alpha$
 there exists a,  parameterized by integer $k$, family $\{\beta_k\}$ of values of $\beta$ satisfying
 relations  (9) or (10). For $\omega=0$ and
 $(\varphi_{\!{}_1},\varphi_{\!{}_2})=(0,0)$, problem reduces to
 that analyzed in Facts 1-2. For $\omega=0$ and
 $(\varphi_{\!{}_1},\varphi_{\!{}_2})=(\pi,\pi)$, it reduces to
 the case described by Facts 3-4.

\begin{figure}[h]
\noindent\rule{15.4cm}{0.4pt}
\begin{center}
{\footnotesize{a)}}$\!\!\!$
\includegraphics[width=5.5cm]{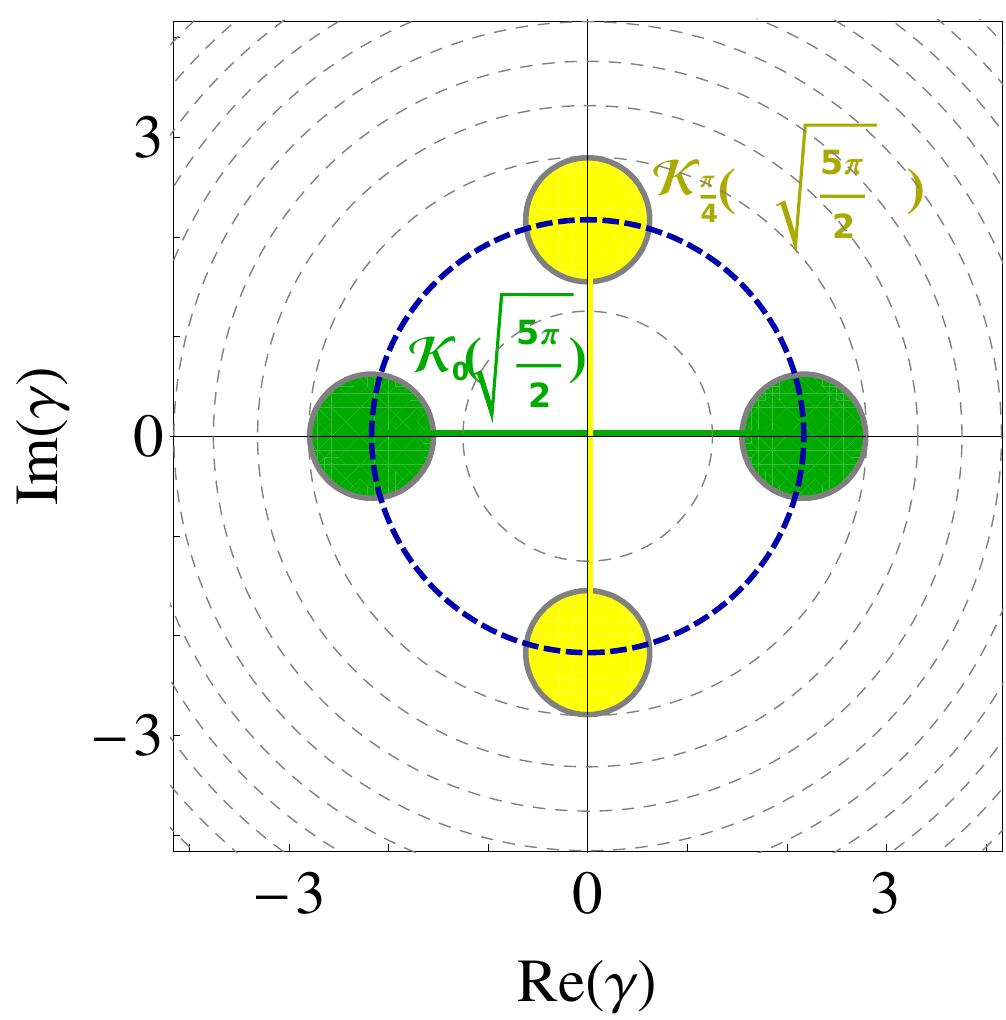} $\qquad${\footnotesize{b)}}$\!\!$ \includegraphics[width=5.5cm]{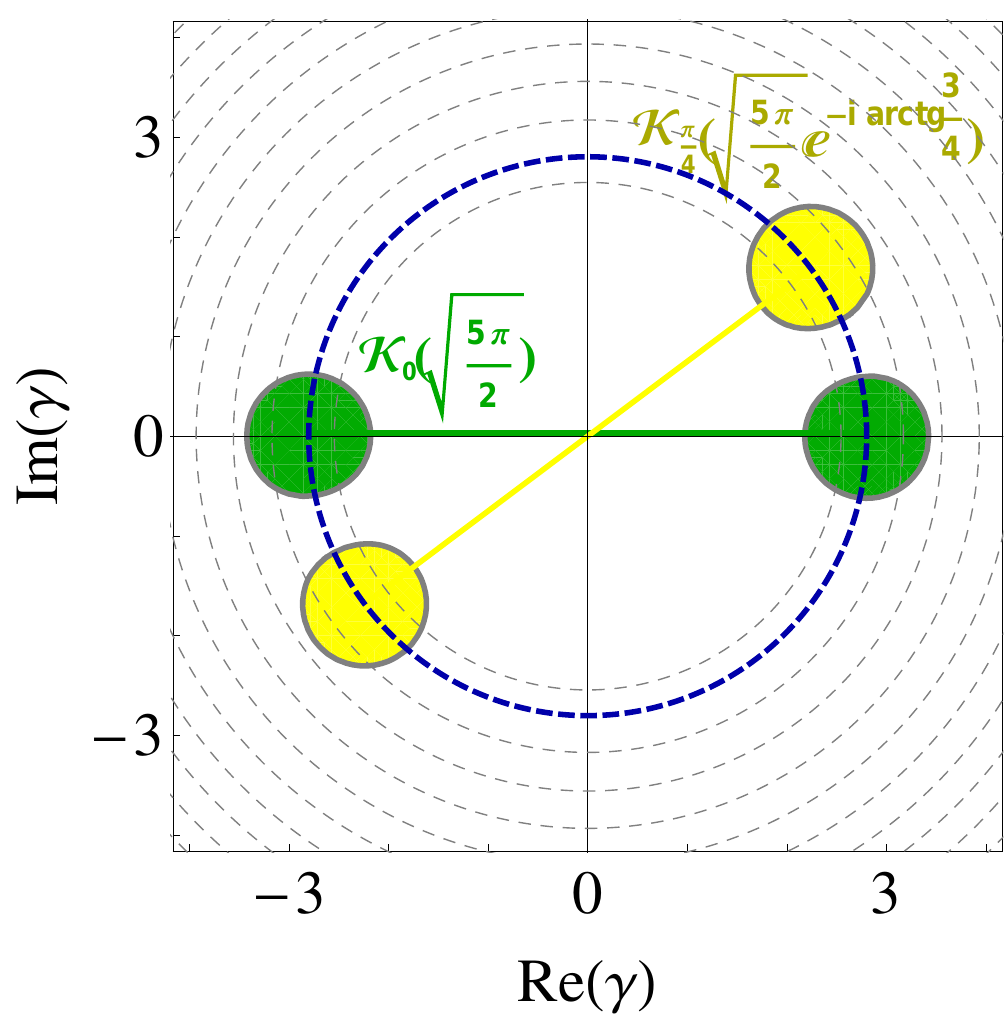}
\end{center}
\caption{\small Phase space representation of orthogonal
vectors ${\mathcal{K}}_{\scriptscriptstyle{0}}(\alpha)$ (green)
and ${\mathcal{K}}_{\varphi_{\!{}_2}}(\beta)$ (yellow) with equal
mean number of photons $|\alpha|^2=|\beta|^2=\sqrt{\omega^2 +
 (1/2+k)^2\pi^2}$, calculated for  $k=1$ and a)~$\omega=0$;
 b)~$\omega=2\pi$. Dashed circles denote the only possible values
 fulfilling  conditions  $\lan{\mathcal{K}}_{\varphi_{\!{}_1}}(\alpha)|{\mathcal{K}}_{\varphi_{\!{}_2}}(\beta)
 \ran=0$ and
 $|\alpha|=|\beta|$,   for given $\omega$. Width of the bands between subsequent
 circles strongly depend on~$\omega$: in case a) corresponding to odd
 cat-like superpositions, described in Section 2.2,  areas of subsequent bands are
always equal to $\pi^2$. In case b) areas of the bands depend on
$k$ and are equal to $\pi^2$ only in the limit of
$|k|\rightarrow\infty$.}\label{fig6}

\noindent\rule{15.4cm}{0.4pt}
\end{figure}

 If we are looking for orthogonal vectors
 $ {\mathcal{K}}_{\varphi_{\!{}_1}}\!(\alpha),\,{\mathcal{K}}_{\varphi_{\!{}_2}}\!(\beta)$
  with the same average numbers of photons,  relation
 $|\beta|^2=|\alpha|^2$ has to be satisfied. Depending on quantization condition, it implies
 either $|\alpha|^2=|\beta|^2=\sqrt{\omega^2 +
 k^2\pi^2}\,$  or
 \ba
 |\alpha|^2=|\beta|^2=\sqrt{\omega^2 +
 (1/2+k)^2\pi^2}. \label{e13}
\ea
 Examples of orthogonal vectors ${\mathcal{K}}_{\varphi_{\!{}_1}}(\alpha)
 $,
 ${\mathcal{K}}_{\varphi_{\!{}_2}}\!(\beta) $  fulfilling (\ref{11B})
 are presented in Figs.~\ref{fig6}.a-b. Both plots were made under
 assumption that
 $\varphi_{\scriptscriptstyle{1}}=0$ and
 {\mbox{$k=1$}}. Vector ${\mathcal{K}}_0(\alpha) $ is plotted in green,
 the orthogonal vector ${\mathcal{K}}_{\varphi_{\!{}_2}}\!(\beta) $ in
yellow. Dashed circles denote the only possible values of $\alpha$
 fulfilling  condition (\ref{e13}) for a set parameter $\omega$.
Figure~\ref{fig6}.a corresponds to $\omega=0$ (for details on
effects of vanishing real part of $\alpha\beta^*$, see paragraph
below); Fig.~\ref{fig6}.b corresponds to $\omega=2\pi$. It is
seen, that non-zero values of $\omega$  modify an angle of phase
space rotation between the orthogonal states, making it
$k$-depended.
 It is also clear that in Fig.~\ref{fig6}.b areas of bands between the
subsequent dashed circles change with $k$, and simple calculation
shows that in the limit of large $|k|$ they approach $\pi^2$. For
$\omega=0$, Fig.~\ref{fig6}.a,  areas of the bands
 are always equal to $\pi^2$, as already mentioned in Section 2.2, where
 case of odd cat-like states was discussed. Presented in that Section, Fig.~\ref{fig2}
 shows an
 example of orthogonal even cat-like superpositions with the same amplitudes, vanishing real part of $\alpha\beta^*$
 ($\omega=0$), and $k=4$.

\paragraph{When real part of $\alpha\beta^*$ vanishes}
Consider a special case when
{\mbox{$\omega\!=\!\mathrm{Re}(\alpha\beta^*)\!=\!0$}}. If
$\cos[2\,\mathrm{Im}(\alpha\beta^*) ]=1$, from (\ref{10A}) follows
that one of the phases  $\varphi_{\!{}_1}$, $\varphi_{\!{}_2}$ has
to be equal to $\pi$, the other can be arbitrary. If
$\cos[2\,\mathrm{Im}(\alpha\beta^*)]=-1$,  one of the phases has
to be equal to $0$ and the other is arbitrary. In
Fig.~\ref{rysTorus} the former case is denoted by the  dashed
white lines,  the later by dashed blue lines, i.e. edges of a
larger yellow-green square. Note, that the case when
$(\varphi_{\!{}_1},\varphi_{\!{}_2})\in[1pt]
\{(0,\pi),(\pi,0) \} $ takes us once again to the orthogonality between odd and even cats.
It is clear that in this case, for a known $\alpha\neq 0$  and one
and only one of the phases equal to 0 or $\pi$, value of $\beta$
is determined unambiguously, while the second phase is arbitrary.

From now on we will assume that real part of $\alpha\beta^*$ is
nonzero, which means that we will consider only phases depicted in
Fig.~\ref{rysTorus} by green or yellow open triangles (without
edges).

\subsubsection{Set values of  $\varphi_{\!{}_1}$, $\varphi_{\!{}_2}$ and $\alpha$}
To avoid repetitions,  we assume now that
{\mbox{$\cos({\scriptstyle{\frac{\varphi_{\!{}_2}+\varphi_{\!{}_1}}{2}
}})
\cos({\scriptstyle{\frac{\varphi_{\!{}_2}-\varphi_{\!{}_1}}{2}
}})\neq 0\,$}} and that $\varphi_{\!{}_1},\varphi_{\!{}_2}$  are
different then~0 or~$\pi$. 
For given phases $\varphi_{\!{}_1}$, $\varphi_{\!{}_2}$,
quantization condition depends on whether point
$(\varphi_{\!{}_1},\varphi_{\!{}_2})$ belongs to a green or yellow
areas in Fig.~\ref{rysTorus} . In the first case, phases are such
that {\mbox{${\cos(\frac{\varphi_{\!{}_1} -
\varphi_{\!{}_2}}{2})}/{\cos(\frac{\varphi_{\!{}_1} + \varphi_{\!{}_2}}{2})} < 0$}},
which imposes condition (\ref{10A}) and, consequently,
(\ref{10B}).
 The second case, when
{\mbox{${\cos(\frac{\varphi_{\!{}_1} -
\varphi_{\!{}_2}}{2})}/{\cos(\frac{\varphi_{\!{}_1} + \varphi_{\!{}_2}}{2})} > 0$}} leads to
conditions (\ref{11A}) and  (\ref{11B}).
 In both cases, for a known
$\varphi_{\!{}_1}$, $\varphi_{\!{}_2}$ and $\alpha\neq 0$, value
of $\beta$ can be determined from (\ref{e12}).

 Table~1 summarizes the results obtained so far:\\

\begin{tabular}{|rl|}
  \hline
 \multicolumn{2}{|c|}{{\bf{Table 1.}} Values of $\beta$ such that
 $\lan{\mathcal{K}}_{\varphi_{\!{}_2}}\!(\beta)|{\mathcal{K}}_{\varphi_{\!{}_1}}\!(\alpha)\ran=0$ for
 given
 $\alpha,\varphi_{\!{}_1}$, $\varphi_{\!{}_2}$.} \\
  \hline
$(\varphi_{\!{}_1}$, $\varphi_{\!{}_2})\in[1pt]$ (green surface in
Fig.~\ref{rysTorus}) & $\Rightarrow\,\beta_k= (\omega - i
k\pi)/\alpha^{*},\;\;\;k\in[1pt]\mathbb{Z}$ \\
  $(\varphi_{\!{}_1}$, $\varphi_{\!{}_2})\in[1pt]$
  (yellow surface  in Fig.~\ref{rysTorus}) & $\Rightarrow\,\beta_k= (\omega - i(k
+ 1/2)\pi)/\alpha^{*},\;\;k\in[1pt]\mathbb{Z}$ \\
 $(\varphi_{\!{}_1}$, $\varphi_{\!{}_2})\in[1pt]$ (red dots in Fig.~\ref{rysTorus})
 & $\Rightarrow\,$ arbitrary $\beta$ \\
  $(\varphi_{\!{}_1}$, $\varphi_{\!{}_2})\in[1pt]$ (black segments in Fig.~\ref{rysTorus})
 & $\Rightarrow\,$ no solutions \\
  \hline
\end{tabular}

\subsubsection{Set values of $\alpha$, $\beta$ and $\varphi_{\!{}_1}$}
To analyze conditions  (\ref{10A}), (\ref{11A}),  for set values
of $\alpha$, $\beta$ and $\varphi_{\!{}_1}$ or set $\alpha$,
$\beta$ and $\varphi_{\!{}_2}$, it is convenient to introduce new
variables:
{\mbox{$a=\mathrm{tg}\big(\frac{\varphi_{\!{}_1}}{4}\big)$,}}
{\mbox{$b=\mathrm{tg}\big(\frac{\varphi_{\!{}_2}}{4}\big)$}}, that
 transform (\ref{10A})-(\ref{11A}) into analytically
solvable quadratic equations. After some calculations it can be
shown
 that, for given $\alpha$, $\beta$ and $\varphi_{\!{}_1}$,
solutions of (\ref{11A}) for $\varphi_{\!{}_2}$ are
\begin{equation}
\begin{aligned}
& \textstyle
\varphi_{\!{}_2}^{+} =
\arctan\!\left[\frac{\sqrt{W(\varphi_{\!{}_1},\alpha,
\beta)^{2} + U(\varphi_{\!{}_1},\alpha, \beta)^{2}} - W(\varphi_{\!{}_1},\alpha,
\beta)}{U(\varphi_{\!{}_1},\alpha, \beta)} \right]\;
\mathrm{ if }\; \mathrm{Re}(\alpha\beta^{\ast})>0, \\
&\textstyle
\varphi_{\!{}_2}^{-} =
\arctan\!\left[
\frac{\sqrt{W(\varphi_{\!{}_1},\alpha,
\beta)^{2} + U(\varphi_{\!{}_1},\alpha,
\beta)^{2}} + W(\varphi_{\!{}_1},\alpha, \beta)}{-U(\varphi_{\!{}_1},\alpha, \beta)}
\right]\;
 \mathrm{if} \; \mathrm{Re}(\alpha\beta^{\ast})<0,
\end{aligned}\label{j15}
\end{equation}
\begin{equation}
\begin{aligned}
\mathrm{where}\qquad\qquad & W(\varphi_{\!{}_1},\alpha, \beta) :=
2\tan({\textstyle{\frac{\varphi_{\!{}_1}}{4}}})\big[ 1 +
\exp(2\mathrm{Re}(\alpha\beta^{\ast}))\big],\\
& U(\varphi_{\!{}_1},\alpha, \beta) := \big[1 - \exp(2
\mathrm{Re}(\alpha\beta^{\ast}))\big]\big[(\tan({\textstyle{\frac{\varphi_{\!{}_1}}{4}}}))^{2}
- 1\big].
\end{aligned}\label{j16}
\end{equation}
Analogously, after introducing
\begin{equation}
\begin{aligned}
&W'(\varphi_{\!{}_1},\alpha, \beta) :=
2\tan({\textstyle{\frac{\varphi_{\!{}_1}}{4}}})\big[ 1 - \exp(2\mathrm{Re}(\alpha\beta^{\ast}))\big],\\
& U'(\varphi_{\!{}_1},\alpha,
\beta) := \big[1 + \exp(2
\mathrm{Re}(\alpha\beta^{\ast}))\big]\big[(\tan({\textstyle{\frac{\varphi_{\!{}_1}}{4}}}))^{2} -
1\big],
\end{aligned}\label{j17}
\end{equation}
solution of (\ref{10A}) can be written as
\begin{equation}
\begin{aligned}
&\textstyle
\varphi_{\!{}_2}'^{+} =
\arctan\!\left[\frac{\sqrt{W'(\varphi_{\!{}_1},\alpha,
\beta)^{2} + U'(\varphi_{\!{}_1},\alpha, \beta)^{2}} - W'(\varphi_{\!{}_1},\alpha,
\beta)}{U'(\varphi_{\!{}_1},\alpha, \beta)} \right]\;
\mathrm{ for }\; \mathrm{Re}(\alpha\beta^{\ast})>0, \\
&\textstyle
\varphi_{\!{}_2}'^{-} = \arctan\!\left[
\frac{\sqrt{W'(\varphi_{\!{}_1},\alpha,
\beta)^{2} + U'(\varphi_{\!{}_1},\alpha,
\beta)^{2}} + W'(\varphi_{\!{}_1},\alpha, \beta)}{-U'(\varphi_{\!{}_1},\alpha, \beta)}
\right]\;
\mathrm{ for }\; \mathrm{Re}(\alpha\beta^{\ast})<0.
\end{aligned}\label{j18}
\end{equation}

We have shown unambiguous solutions of (\ref{scal}) in the case of
known parameters $\alpha$, $\beta$ and~$\varphi_{\!{}_1}$. In the
case when~$\varphi_{\!{}_1}$ is a variable and~$\varphi_{\!{}_2}$
a known parameter, values of~$\varphi_{\!{}_1}$ can be find in the
same fashion because all the equations used were symmetric under
transformation $\varphi_{\!{}_1}\leftrightarrow
\varphi_{\!{}_2}$. This ends analysis of equation (\ref{scal}).

To summarize results of this subsection: If {\mbox{$\cos[2\,
\mathrm{Im}(\alpha\beta^{\ast})] \!= \!\pm 1$}} one can always
find phases $\varphi_{\!{}_1}, \varphi_{\!{}_2}$ such that
$|{\mathcal{K}}_{\varphi_{\!{}_2}}\!(\beta)\ran$ is orthogonal to
$|{\mathcal{K}}_{\varphi_{\!{}_1}}\!(\alpha)\ran $. Moreover,
\begin{itemize}
\item{for $\mathrm{Re}(\alpha\beta^{\ast}) >0$
and any $\varphi_{\!{}_1}$ different then 0 or $\pi$ there exists
exactly one $\varphi_{\!{}_2} $ that makes respective vectors
orthogonal. Both $\varphi_{\!{}_1} $ and $\varphi_{\!{}_2} $ are
simultaneously smaller or larger then $\pi$.  }
\item{for $\mathrm{Re}(\alpha\beta^{\ast}) <0$ and  any $\varphi_{\!{}_1}$ different then 0 or $\pi$
there exists  exactly one $\varphi_{\!{}_2} $ that makes
respective vectors orthogonal, and either $\varphi_{\!{}_1} $ or
$\varphi_{\!{}_2} $ is larger then $\pi$.  }
\item{for $\mathrm{Re}(\alpha\beta^{\ast}) =0$ and  $ \cos[2\, \mathrm{Im}(\alpha\beta^{\ast})] =  1$,
to obtain orthogonality one phase has to be equal to $\pi$, the
other phase can be arbitrary (white dashed lines in
Fig.~\ref{rysTorus}). Special case when
$(\varphi_{\!{}_1},\varphi_{\!{}_2})\in[1pt]
\{(0,\pi),(\pi,0) \} $  corresponds to the always
orthogonal cat-like superpositions of different parity.}
\item{for $\mathrm{Re}(\alpha\beta^{\ast}) =0$ and  $\cos[2\, \mathrm{Im}(\alpha\beta^{\ast})] = - 1$,
 one of the
phases has to be equal to $0$, second can be arbitrary (dashed
blue lines in Fig.~\ref{rysTorus}). Special case,
$(\varphi_{\!{}_1},\varphi_{\!{}_2})\in[1pt]
\{(0,\pi),(\pi,0) \}, $  corresponds to the always
orthogonal cat-like superpositions of different parity.}
\end{itemize}

\section{Summary and outlook}

We have presented a full analytical solution of a problem of
finding orthogonal vectors within a set of cat-like superpositions
of coherent states, (\ref{e1}). We have
 shown that in the case of known $\alpha, \beta $, and $\varphi_{\!{}_1}$ the condition
 $\lan{\mathcal{K}}_{\varphi_{\!{}_2}}
 (\beta)|
{\mathcal{K}}_{\varphi_{\!{}_1}}(\alpha)\ran=0\,$ determines
$\varphi_{\!{}_2}$ unambiguously, thus, measurements of the scalar
product can be used to measure the phase.
 We have also shown that for any given cat-like superposition
${\mathcal{K}}_{\varphi_{\!{}_1}}\!(\alpha)$ and set
$\varphi_{\!{}_2}$, such that
$(\varphi_{\!{}_1},\varphi_{\!{}_2})$ belongs to either green or
yellow areas in Fig.~\ref{rysTorus}, there exist a whole family of
vectors of the form ${\mathcal{K}}_{\varphi_{\!{}_2}}\!(\beta)$
orthogonal to ${\mathcal{K}}_{\varphi_{\!{}_1}}\!(\alpha)$, and
have presented an explicit solutions. We have proved that the
considered orthogonality condition imposes on antisymmetric
(simplectic) form $\mathrm{ Im}(\alpha\beta^*) $ a quantization
condition permitting only discreet values, and has no such
restriction on metric form $\mathrm{Re}(\alpha\beta^*) $.

Results presented in this paper show, among others, that
structures build from superposition of coherent states let 
directly deterministically  distinguish between them -- in
contrast to only probabilistic distinguisability between `single'
coherent states.
 This fact has potentially many
applications:  from ability to perfectly distinguish different cat
states follows, in principle, possibility to use precise
measurement of a scalar product for a precise measurement of
phase, and vice versa. In the context of quantum communication,
existence of infinite sets of orthogonal states allows to send
binary sequences without repetition of code words,  and a
condition $
\cos[2\,
\mathrm{Im}(\alpha\beta^{\ast})] =  \pm 1$ adds a possibility of
additional spin-like encoding.

{\bf Acknowledgments}  L.P. thanks Prof. Ray-Kuang Lee for his
hospitality and stimulating discussions. This work was supported
by NTHU project no. 104N1807E1.

\section*{Appendix A:  Notation}
In this paper a standard notation was used: Greek letters
$\alpha,\beta,\gamma...$ denote complex numbers and
$|\alpha\ran,|\beta\ran,|\gamma\ran$ are the corresponding
coherent states. Set of real and complex numbers are denoted by
$\mathbb{R}$ and $\mathbb{C}$, respectively, while
$\mathbb{R}^\times$ and $\mathbb{C}^\times$ denote sets of real or
complex numbers without zero. A set of integer numbers is denoted
as $\mathbb{Z}$, and a set of natural numbers (with zero) as
$\mathbb{N}$, whereas $\mathbb{N}^\times
\defeq\mathbb{N}\setminus[1pt]
\{0\} $.

\section*{Appendix B:  }
Decomposition of a complex number $\gamma\in[1pt]\mathbb{ C}$ into
real and imaginary parts, $\mathrm{Re}(\gamma)$ and $
\mathrm{Im}(\gamma)$, defines a canonical isomorphism of real
vector spaces
\ba
\vec{{}}\,:\mathbb{C}\ni[1pt]\gamma\rightarrow
\vec{\gamma}\defeq \left(\!\!\!\begin{array}{c}
\mathrm{Re}(\gamma)\\\mathrm{Im}(\gamma)\end{array}
\!\!\!\right)
\in[1.5pt] \mathbb{R}^2.
\ea
It means that the multiplication of complex numbers
$\alpha\beta^*$ defines two bilinear forms on $\mathbb{R}^2$
\ba
&&g:\mathbb{R}^2\!\ttimes[1pt]\mathbb{R}^2\ni[1pt](\vec{\alpha},\vec{\beta})
\rightarrow g(\vec{\alpha},\vec{\beta})\defeq\mathrm{
Re}(\alpha\beta^*)\in[1pt]\mathbb{ R} \\
&&h:\mathbb{R}^2\!\ttimes[1pt]\mathbb{R}^2\ni[1pt](\vec{\alpha},\vec{\beta})
\rightarrow  h(\vec{\alpha},\vec{\beta})\defeq\mathrm{
Im}(\alpha\beta^*)\in[1pt]\mathbb{ R}.
\ea
Form $g$ is symmetric and defines a metric (Riemann) structure on
$\mathbb{R}^2$. Form $h$ is antisymmetric and defines a symplectic
structure on $\mathbb{R}^2$.

Results of Section 3.2  show that orthogonality condition between
vectors ${\mathcal{K}}_{\varphi_{\!{}_1}}\!(\alpha)$ and
${\mathcal{K}}_{\varphi_{\!{}_2}}\!(\beta)$ impose quantization
condition on possible values of symplectic form $
h(\vec{\alpha},\vec{\beta})$ on $\mathbb{R}^2$,  as it is clearly
seen from (\ref{10B}) and (\ref{11B}). At the same time connected
with phases $\varphi_{\!{}_1},\varphi_{\!{}_2}$ values of metric
form $g(\vec{\alpha},\vec{\beta})$ can take on arbitrary real
values and are continuous (see (\ref{10A}) and (\ref{11A})).


\bibliography{Catbibfile}

\end{document}